\documentclass[prb,twocolumn,showpacs,preprintnumbers,amsmath,amssymb,superscriptaddress]{revtex4-1}

\usepackage{graphicx}% Include figure files
\usepackage{dcolumn}% Align table columns on decimal point
\usepackage{bm}% bold math
\newcommand{\bfpsi}{\mbox{\boldmath$\psi$}}

\begin{document}

\title{Thermal evolution of quasi-1D spin correlations within the anisotropic triangular lattice of $\alpha$-NaMnO$_2$}% Force line breaks with \\

\author{Rebecca L. Dally}
\affiliation{Materials Department, University of California, Santa Barbara, California 93106-5050, USA}
\author{Robin Chisnell}
\affiliation{NIST Center for Neutron Research, National Institute of Standards and Technology, Gaithersburg, Maryland 20899, USA}
\author{Leland Harriger}
\affiliation{NIST Center for Neutron Research, National Institute of Standards and Technology, Gaithersburg, Maryland 20899, USA}
\author{Yaohua Liu}
\affiliation{Neutron Scattering Division, Oak Ridge National Laboratory, Oak Ridge, TN 37831, USA}
\author{Jeffrey W. Lynn}
\affiliation{NIST Center for Neutron Research, National Institute of Standards and Technology, Gaithersburg, Maryland 20899, USA}
\author{Stephen D. Wilson}
\email{stephendwilson@ucsb.edu}
\affiliation{Materials Department, University of California, Santa Barbara, California 93106-5050, USA}
\date{\today}% It is always \today, today,

\begin{abstract}
Magnetic order on the spatially anisotropic triangular lattice of $\alpha$-NaMnO$_2$ is studied via neutron diffraction measurements. The transition into a commensurate, collinear antiferromagnetic ground state with $\mathbf{k}=(0.5, 0.5, 0)$ was found to occur below $T_\mathrm{N}=22$ K.  Above this temperature, the transition is preceded by the formation of a coexisting, short-range ordered, incommensurate state below $T_\mathrm{{IC}}=45$ K whose two dimensional propagation vector evolves toward $\mathbf{k}=(0.5, 0.5)$ as the temperature approaches $T_{\mathrm{N}}$.  At high temperatures ($T>T_\mathrm{{IC}}$), quasielastic scattering reveals one-dimensional spin correlations along the nearest neighbor Mn-Mn ``chain direction" of the MnO$_6$ planes.  Our data are consistent with the predictions of a mean field model of Ising-like spins on an anisotropic triangular lattice, as well as the predominantly one-dimensional Heisenberg spin Hamiltonian reported for this material.      
\end{abstract}

%\pacs{61.72.Nn, 71.70.Ej, 75.25.-j, 75.30.Gw, 75.50.Ee, 77.80.B-, 83.85.Hf}

\maketitle

%\tableofcontents

\section{Introduction}
Two-dimensional triangular lattice antiferromagnets are a canonical example of geometrically frustrated spins. A wide variety of unconventional magnetic states are realized in these systems as the magnetic ions are populated with either classical,\cite{Kawamura_1984} highly quantum (i.e.\ $S=1/2$), \cite{Huse_1988, Jolicoeur_1989, Singh_1992, Bernu_1994} or spin-orbit entangled moments.\cite{Shirata_2012, Kimchi_2014, Ranjith_2017, Li_2018}  Furthermore, by perturbing the underlying lattice and the mode through which the geometric frustration is lifted, additional new magnetic phase behaviors may be realized. Specifically, in reducing the degeneracy of competing magnetic exchange pathways in this model by one, a spatially anisotropic triangular lattice is generated. This has the potential of stabilizing predominantly one-dimensional magnetic interactions within the underlying spin system. 

The anisotropic triangular lattice can be viewed as interconnected isosceles triangles with two long legs and one short leg, forming ``chains" of nearest neighbors. Each atom has four next-nearest neighbors, which frustrates antiferromagnetic (AF) ordering. The relative strength of exchange interactions between nearest neighbors ($J_1$) to that of next-nearest neighbors ($J_2$), as well as inherent exchange and single-ion anisotropies, determines the effective dimensionality of the spin system and whether it can realize long-range order.

In the case of $S=1/2$ moments decorating the lattice, spin liquid,\cite{Yunoki_PRB_2006} collinear antiferromagnetic, and dimer orders \cite{Starykh_PRL_2007} are all predicted depending on the extent of the exchange anisotropy. For the case of $S=1$ moments on the lattice, as the effective interchain frustration $J_2\rightarrow 0$, the spin system is driven toward the limit of isolated antiferromagnetic spin chains where the ground state is a Haldane singlet.\cite{Haldane_1983} For the case of further increased $S$ toward more conventional moments, when $0.7\lesssim J_2/J_1\lesssim 1$, the ground state remains that of the isotropic triangular lattice ($120^{\circ}$ spiral spin state),\cite{Nakano_JPSJ_2013} and for lesser values of $J_2/J_1$, N{\'e}el and helical order are known to stabilize.\cite{Li_EPJB_2012} Parameters such as spin-orbit effects or Dzyaloshinskii-Moriya interactions can further tune the system into different regions of phase space. 

In this paper, we utilize neutron scattering techniques to explore the magnetic phase behavior in single crystals of an $S=2$ anisotropic triangular spin system, $\alpha$-NaMnO$_2$, with particular focus given to how the spins evolve toward their collinear AF ground state. We show that the transition from the high temperature paramagnetic phase into the low temperature N{\'e}el state is characterized by three temperature regimes. The first is a high temperature regime between $T_{1\mathrm{D}}\approx 200$ K $>T>T_{\mathrm{IC}}\approx 45$ K, where the nearest neighbor intrachain interaction $J_1$ and frustrated interchain interactions drive quasi-one-dimensional (1D) correlations along the short $b$-axis (the Mn-Mn nearest-neighbor chain direction).  As the material is cooled below the expected N{\'e}el temperature of 45 K, a phase comprised of competing short-range incommensurately modulated and short-range commensurate AF states appears. Below 22 K, the commensurate AF ground state dominates with correlation lengths along the interchain and interplane axes truncated by the microstructure of the crystal. The formation of an intermediate, incommensurately modulated state upon cooling agrees with the expectations of a mean-field model of Ising moments on an anisotropic triangular lattice. Additionally, the persistence of quasi-1D spin correlations at temperatures far above the AF ordered state is consistent with the inherent one dimensionality of the spin dynamics in this compound. 

\section{Experimental methods}

The single crystals used throughout this study were synthesized using the floating zone method with details of the procedure and characterization found in Ref.~\onlinecite{Dally_JCrystGrowth}. The samples grown under these conditions were shown, via NMR, to have a minimal amount of $\beta$-phase polymorph ($<1$\%) inclusions and a low percentage of stacking faults (4\%). Inductively coupled plasma  atomic emission spectroscopy (ICP-AES) measurements yielded a Na:Mn molar ratio of 0.90:1.00; however, as described in the appendix of this paper, this ratio includes an impurity phase of Mn$_3$O$_4$ coherently grown within the lattice.  Taking this impurity phase into account, the nominal stoichiometry of the $\alpha$-NaMnO$_2$ phase is Na$_{0.98}$MnO$_2$. For the remainder of this manuscript, we refer to the material as $\alpha$-NaMnO$_2$.  Magnetization measurements were taken using a Quantum Design PPMS Vibrating Sample Magnetometer (VSM) with the field oriented $\mathbf{H} \parallel [0, 1, 0]$ of $\alpha$-NaMnO$_2$.

Neutron powder diffraction data were collected using the BT-1 neutron powder diffractometer at the NIST Center for Neutron Research.  For these measurements, a 3 g single crystal of $\alpha$-NaMnO$_2$ was ground and sealed in a vanadium container inside a dry He-filled glove box, which was then placed in a closed cycle refrigerator (CCR). A Cu(311) monochromator with a $90^{\circ}$ take-off angle, $\lambda$ = 1.5397(2) {\AA}, and in-pile collimation of $60^\prime$ were used. Data were collected over the range of $2\theta =3-168^{\circ}$ with a step size of $0.05^{\circ}$. Analysis of powder data was performed using the FullProf \cite{FullProf} and FAULTS \cite{FAULTS} analysis suites for Rietveld refinement and SARAh \cite{SARAh} for representational analysis. 

Neutron time-of-flight measurements were taken on the CORELLI instrument\cite{Rosenkranz2008} at the Spallation Neutron Source at Oak Ridge National Laboratory. A 0.5 g single crystal was mounted on a thin aluminum plate and sealed in an aluminum can under an inert environment, which was then placed in a CCR. The CORELLI beam was modulated by a stochastic chopper where, using a cross-correlation method, the quasielastic signal was extracted.

High resolution, triple-axis neutron scattering measurements were performed using the cold neutron spectrometer SPINS, at the NIST Center for Neutron Research. A flat analyzer and $E_{\mathrm{i}}=5$ meV were used, and unless otherwise specified, data taken in the $(H,K,0)$ scattering plane used the collimation, open$-40^\prime-40^\prime-$open, denoting the collimations before the monochromator, sample, analyzer, and detector positions, respectively. Data taken in the $(H,H,L)$ scattering plane used open$-80^\prime-80^\prime-$open collimation. All triple-axis data were analyzed with fits to Voigt functions, where the Gaussian width was fixed as the spectrometer resolution modified by the inclusion of a sample mosaic of $20^\prime$. The convolved widths were calculated using the program ResLib. \cite{ResLib} Throughout the manuscript, the $\mathbf{Q}$ scattering vector is given in reciprocal lattice units $(H,K,L)$, where $\mathbf{Q}\left[ \mathrm{\AA} ^{-1} \right] =2\pi \left(\frac{H}{a\sin \beta}, \frac{K}{b}, \frac{L}{c\sin \beta} \right)$.

\begin{figure}[t]
\includegraphics[scale=0.38]{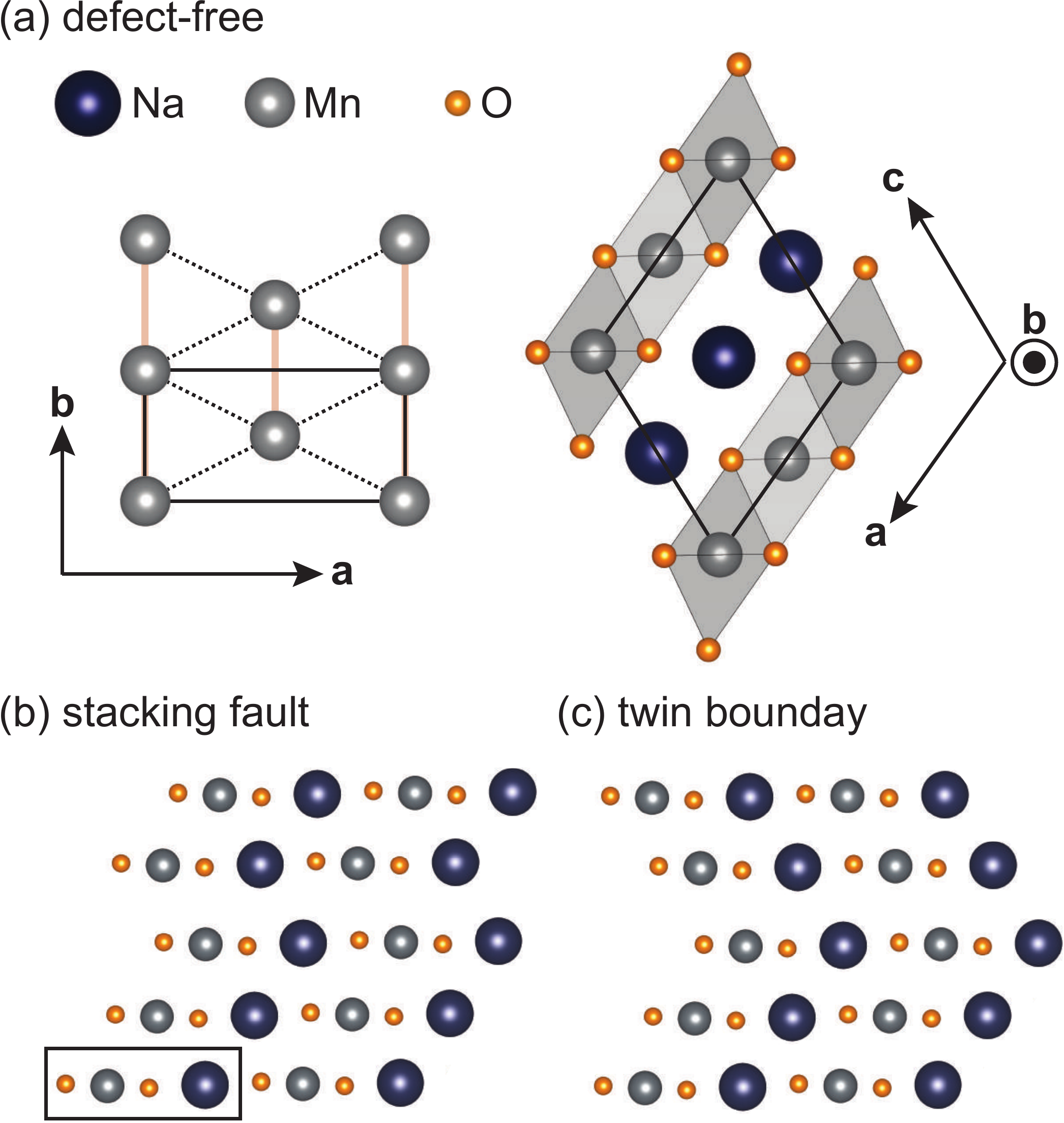}
\caption{\label{fig:structure}(Color online) Summary of the $\alpha$-NaMnO$_2$ structure and defect types. (a) The defect-free \textit{ab}- and \textit{ac}-planes, with the chemical unit cells for both outlined in black. The \textit{ab}-plane schematic shows a single layer of Mn atoms with nearest neighbors connected by solid orange lines and next-nearest neighbors connected by dashed black lines. MnO$_6$ octahedra are shaded in gray in the \textit{ac}-plane schematic. The orientation of the \textit{ac}-plane was made to highlight the direction in which defects form---breaking symmetry along the $\left[ -1,0,1 \right]$ direction---and matches the orientation of the structures in panels (b) and (c). The black rectangle in (b) defines the repeating unit of a layer which can stack to form the defect-free structure in (a), the stacking fault in (b), or the twin boundary in (c).}
\end{figure}

Before discussing the magnetism in $\alpha$-NaMnO$_2$, it is worth reviewing the lattice and complex morphology inherent to crystals of this material.  Specifically, $\alpha$-NaMnO$_2$ crystallizes in the NaFeO$_2$-type structure with a monoclinic $C2/m$ unit cell as shown in Fig.~\ref{fig:structure} (a). The Mn cations are octahedrally coordinated by oxygens, forming MnO$_6$ edge-sharing sheets within the $ab$-plane. These MnO$_6$ layers alternate with layers of octahedrally coordinated Na cations along the $c$-axis, forming a distorted O$3$ (O$3'$) stacking structure. As Mn$^{3+}$ is a strongly Jahn-Teller active ion, the MnO$_6$ octahedra are coherently distorted along the $[-1, 0, 1]$ crystallographic direction.

\section{\label{sec:structure}Structure and morphology of $\alpha$-N\lowercase{a}M\lowercase{n}O$_2$}

Jahn-Teller distorted MnO$_6$ octahedra form the basis for the anisotropic triangular lattice of Mn ions and differentiate this material from the $R \overline{3} m$-type lattice of canonical ABO$_2$ isotropic triangular lattice structures.  High spin Mn$^{3+}$, $3d^4$ moments with $S=2$ decorate the MnO$_6$ planes, and nearest neighbor Mn atoms form a dominant AF exchange pathway, $J_1$, along the short $b$-axis.  This enhanced $J_1$ coupling is unfrustrated along this ``chain" axis while the next-nearest neighbor AF exchange remains frustrated by four equivalent $J_2$ exchange pathways.  This remnant, frustrated interchain $J_2$ coupling renormalizes the magnetic interactions in this material to be quasi-1D.\cite{Stock_PRL_2009} 

The single-ion anisotropy inherent to the Mn$^{3+}$ sites is easy-axis such that the energy is minimized when the Mn moments align along the $d_z^2$ orbital, which orients toward the apical oxygens in the Jahn-Teller distorted octahedra. This favors a collinear ground state, which, when coupled with the antiferromagnetic $J_1$ and $J_2$ exchange parameters leads to $\mathbf{k}_1=\left( \frac{1}{2}, \frac{1}{2}, 0 \right)$ ordering. There is also a second propagation vector allowed via symmetry, leading to a second \textit{k}-domain, with $\mathbf{k}_2= \left( -\frac{1}{2}, \frac{1}{2}, 0 \right)$. 

As a further consequence of the large Jahn-Teller distortion of the MnO$_6$ octahedra, the $\alpha$-NaMnO$_2$ lattice has a strong tendency to develop stacking defects. As illustrated in Figs.~\ref{fig:structure} (b) and (c), these are distinct from the conventionally expected faults within the stacking sequence of planar materials, but instead form a fault across the MnO$_6$ layers. In the isolated case, these defects may be thought of as monoclinic twin domain boundaries (Fig.~\ref{fig:structure} (c)). The tendency to fault is high enough that large numbers of consecutive stacking faults may form within the lattice and become an intergrowth of the competing $\beta$-phase polymorph,\cite{Abakumov, Billaud_JACS_2014, Radin_ChemMat_2018} which can be viewed as the densest possible arrangement of stacking faults. Oxygen atoms at fault boundaries need not shift to accommodate the defects, consequently costing little energy. As a result, bulk single crystals of $\alpha$-NaMnO$_2$ are especially prone to these defects, and, while they can be minimized, they remain a source of structural disorder that must be accounted for.

\begin{figure}[t]
\includegraphics[scale=0.29]{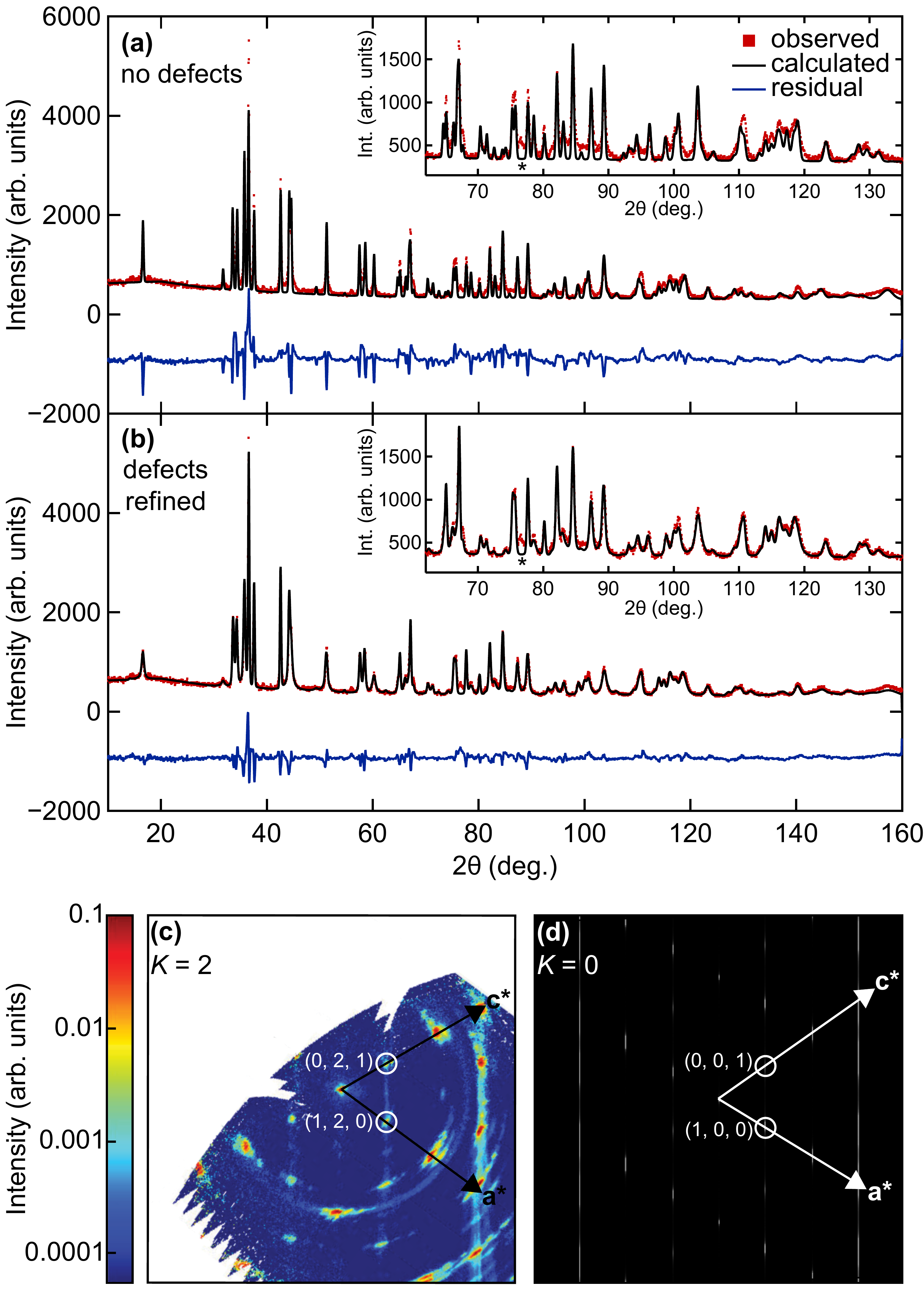}
\caption{\label{fig:NPD_FaultsSim}(Color online) Neutron data and simulations/refinements of the ${\alpha}$-NaMnO$_2$ structure using the program FAULTS. (a) The refinement using a defect-free model as discussed in the text. (b) The refinement allowing the defect percentages to refine. The insets in (a) and (b) show the high $2\theta$ range which highlights the differences between the defect-free fit and the defect-refined fit. The asterisk (*) symbol denotes a Bragg peak associated with Mn$_3$O$_4$ intergrowth. (c) Data from the CORELLI experiment, showing the $(H,2,L)$ scattering plane. (d) A simulation of the $(H,0,L)$ scattering plane using the refined percentage of stacking defects found from the fit presented in (b).}
\end{figure}

\begin{figure*}[t]
\includegraphics[scale=0.20]{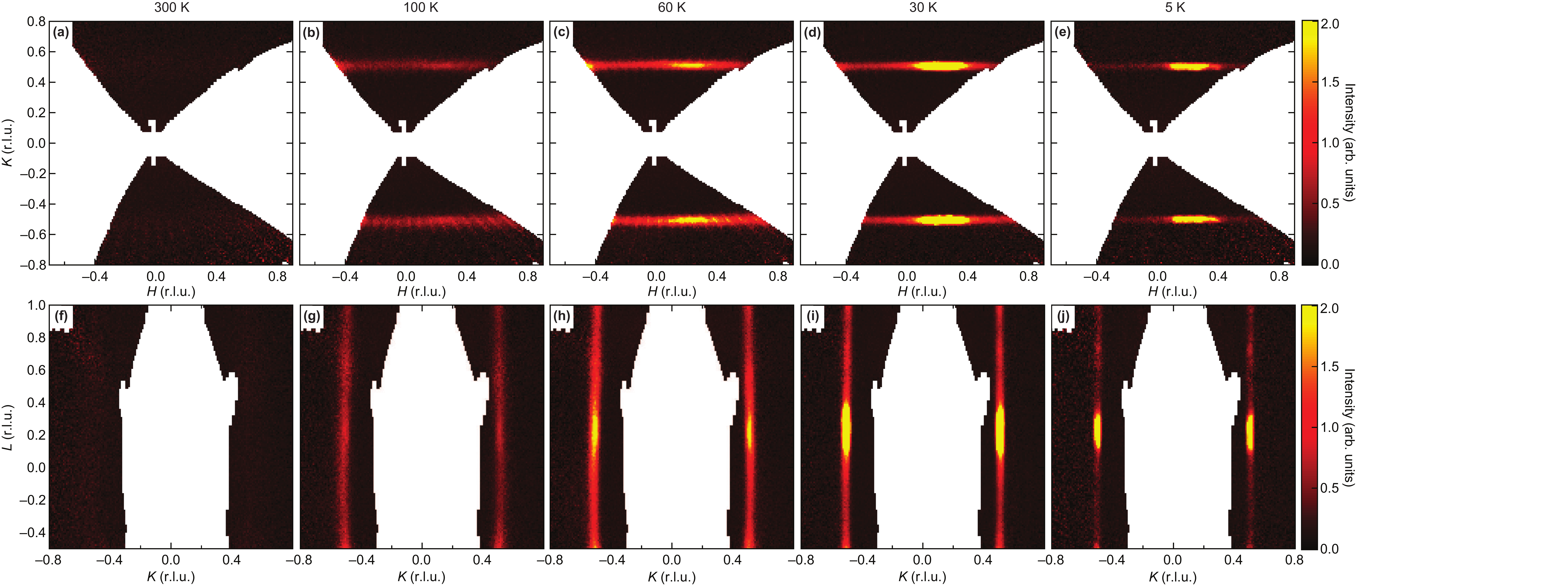}
\caption{\label{fig:CORPLOTS}(Color online) Slices of scattering planes from the neutron spectrometer, CORELLI, at different temperatures. Panels (a)-(e) show slices of the (\textit{H}, \textit{K}, 0.25) plane with 0.15 $<$ \textit{L} (r.l.u.) $<$ 0.35. Panels (f)-(j) show slices of the (0.25, \textit{K}, \textit{L}) plane with 0.15 $<$ \textit{H} (r.l.u.) $<$ 0.35. Maxima along the diffuse scattering rods correspond to a different set of diffuse scattering rods along $[\overline{1}, 0, 1]$, which cut through the planes at these points.}
\end{figure*}

Figs.~\ref{fig:NPD_FaultsSim} (a) and (b) demonstrate the effect that stacking defects have on the intensity distribution in the neutron powder diffraction profile of a crushed single crystal with just a 4.5\% fraction of stacking defects. The program FAULTS was used to refine the models for Figs.~\ref{fig:NPD_FaultsSim} (a) and (b). The refinement for Fig.~\ref{fig:NPD_FaultsSim} (a) allowed the lattice parameters, scale factor, oxygen atom positions and layer-to-layer translation to vary, but the defect percentage was fixed to be zero. Additionally, the pseudo-Voigt line shape parameters were fixed to be the BT-1 instrumental resolution. The peak shape fit is poor, and the high residual reflects the mismatch to the data throughout the entire $2\theta$ range, yielding a $\chi ^2 = 17.3$ and an R-Factor $=13.6$. Alternatively, by allowing the introduction of a refined percentage of stacking defects, the fit vastly improves as shown in Fig.~\ref{fig:NPD_FaultsSim} (b), yielding a $\chi ^2 = 4.4$ and an R-Factor $=6.7$. The optimal fit resulted in a model that within any given layer, there is a 4.5(3)\% chance of a stacking defect forming on the next layer, and after this initial defect, there is a 77.1(3)\% chance of this defect becoming a twin boundary and a 22.9(3)\% chance of the defect forming a stacking fault. Neither the model for Fig.~\ref{fig:NPD_FaultsSim} (a) or (b) included the Mn$_3$O$_4$ impurity, where the only sign of its presence at room temperature in the diffraction pattern is the peak marked with the asterisk (*). 

The redistribution of intensities in the neutron powder diffraction profile due to the stacking defects presents a challenge in refining the magnetic structure via traditional Rietveld methods. Namely, the magnetic Bragg peaks in the diffraction profile are subject to similar deviations in intensity due to the structural faulting.  Nevertheless, an attempt was made to extract quantitative results from $T=5$ K neutron powder diffraction data in order to refine the direction and magnitude of the Mn$^{3+}$ moments. The moments were found to point 38(2)$^{\circ}$ out of the $ab$-plane with a magnitude of 2.42(7) $\mu _{\mathrm{B}}$, and the reduced moment size from the expected 4 $\mu _{\mathrm{B}}$ is consistent with earlier reports.\cite{Giot_2007} The methodology for extracting these results is discussed in more detail in Sect.~\ref{sec:app} of the Appendix. Additionally, the small percentage of Mn$_3$O$_4$ intergrowth and its impact on scattering measurements are discussed further in the Appendix. 

When accounting for both structural and magnetic twinning effects, there are necessarily four magnetic domains---one $\mathbf{k}_1$ and one $\mathbf{k}_2$ domain within each structural twin.  Within the monoclinic unit cell, one can consider for instance the lowest $\mathbf{Q}=(0.5, 0.5, 0)$ AF Bragg reflection and its magnetic twin located at $\mathbf{Q}=(-0.5, 0.5, 0)$.  The structural twin will then also give AF Bragg peaks from each of these magnetic domains close to the $\mathbf{Q}=(0, -0.5, -0.5)$ and $\mathbf{Q}=(0, -0.5, 0.5)$ positions. Naively, within the first Brillouin zone, this preserves the magnetic scattering at an AF Bragg position such as $(0.5, 0.5, 0)$ as originating from a single domain; however, the lattice disorder due to stacking faults also imparts a structurally-driven diffuse component to the magnetic scattering at this position. 

Specifically, quasi-2D rods of scattering arise when the periodic condition for the crystal becomes broken by stacking faults, and in $\alpha$-NaMnO$_2$, the periodicity along the $[\overline{1}, 0, 1]$ direction becomes interrupted. This means that diffuse rods of intensity emanating from crystallographic twins' Bragg peaks will appear along the $[\overline{1}, 0, 1]$ direction in reciprocal space, such as those shown in Figs.~\ref{fig:NPD_FaultsSim} (c) and (d). Fig.~\ref{fig:NPD_FaultsSim} (c) shows the $(H,2,L)$ plane from the CORELLI experiment. Diffuse rods appearing along $L$ due to the stacking defects can be reproduced in a simulation from FAULTS (Fig.~\ref{fig:NPD_FaultsSim} (d)) using the refined percentage of defects found from the neutron powder diffraction refinement. 

The equivalent effect of diffuse scattering rods along the $[\overline{1}, 0, 1]$ direction along will occur for each of the magnetic domains' Bragg peaks as well. These rods can cut through the planes of quasi-1D magnetic correlations and will be referenced later in the manuscript. It should be noted that some of the Bragg peaks in the data and simulation of Figs.~\ref{fig:NPD_FaultsSim} (c) and (d) are not allowed by the $C2/m$ symmetry of pristine $\alpha$-NaMnO$_2$. Specifically, the $(1,2,0)$ position in the data of Fig.~\ref{fig:NPD_FaultsSim} (c), and the $(1,0,0)$ position in the simulation of Fig.~\ref{fig:NPD_FaultsSim} (d), are highlighted, but are the result of crystallographic twins. Additional peaks which cannot be indexed by crystallographic twins can be indexed by the Mn$_3$O$_4$ impurity ($\approx6$\%). Reciprocal space maps showing the location of Mn$_3$O$_4$ scattering with respect to the $\alpha$-NaMnO$_2$ lattice can be found in Fig.~\ref{fig:intergrowthrecip} of the Appendix. 

\section{Single crystal neutron scattering results}

\subsection{High temperature quasi-1D correlations: $200\mathrm{\ K}>T>T_{\mathrm{IC}}$}

\begin{figure*}[t]
\includegraphics[scale=0.25]{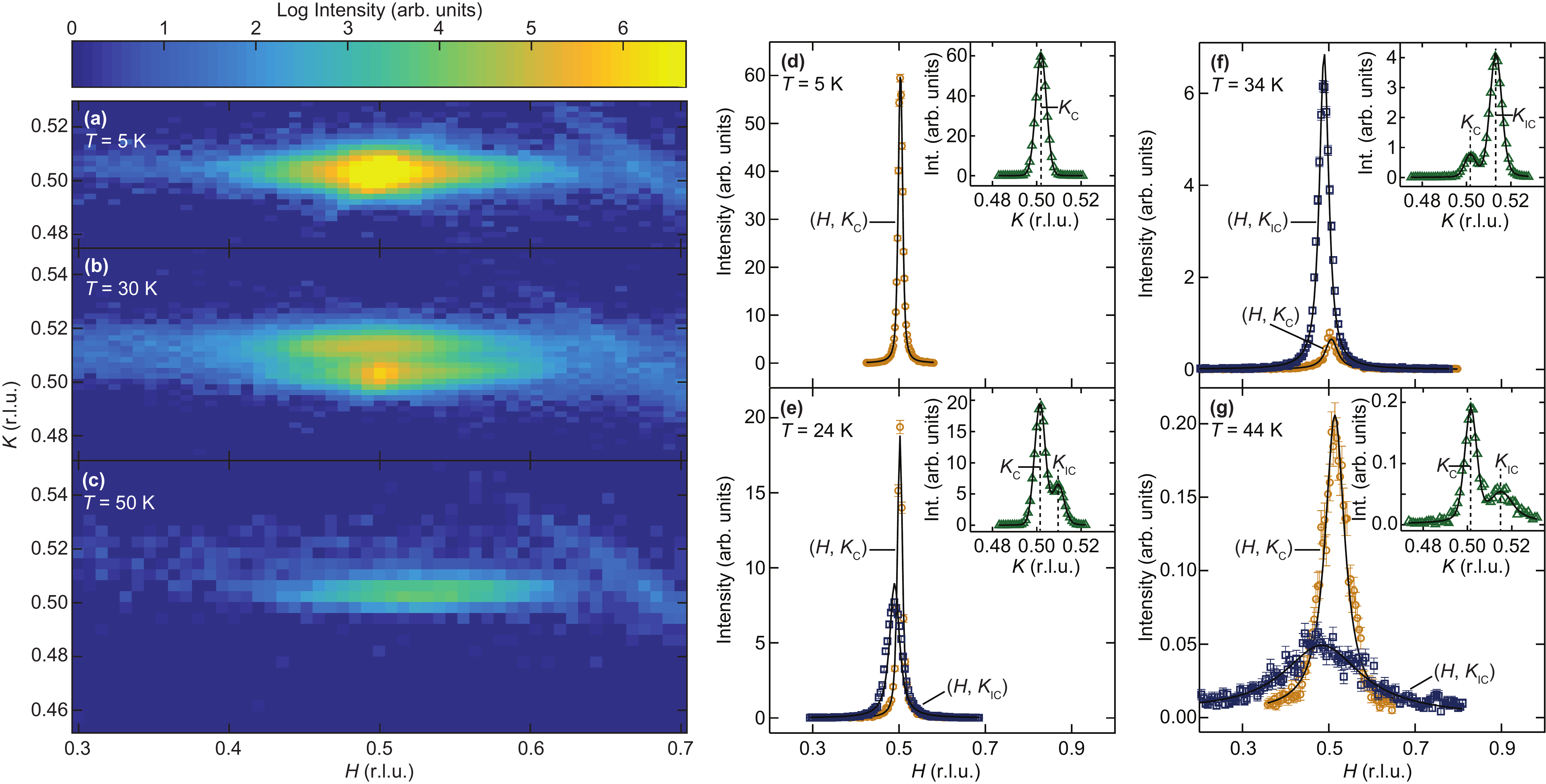}
\caption{\label{fig:meshfits}(Color online) Temperature dependence of the scattering in the $(H,K,0)$ plane about the magnetic ordering wavevector. Panels (a), (b), and (c) were taken at 5 K, 30 K, and 50 K, respectively. (d)-(g) $H$ scans at select temperatures, where $K$ is either at the commensurate position, $K = K_{\mathrm{C}} = 0.5$ r.l.u., or the incommensurate position $K = K_{\mathrm{IC}}(T)$. Both $K_{\mathrm{C}}$ and $K_{\mathrm{IC}}$ are indicated in the insets of each panel, which are $K$ scans taken at $H=0.5$ r.l.u.. Only one peak, located at $K_{\mathrm{C}}$, was observed for the $T=5$ K data shown in panel (d). $L=0$ r.l.u. for all figures, and solid black lines are Voigt function fits.}
\end{figure*}

Short-range magnetic correlations begin to develop in $\alpha$-NaMnO$_2$ well above the nominal ordering temperature. Zero-field cooled susceptibility data previously have shown a broad peak centered around $T_{1\mathrm{D}}=200$ K, characteristic of the onset of 1D magnetic correlations in $\alpha$-NaMnO$_2$. \cite{Giot_2007,Zorko_PRB_2008} To probe the development of spin correlations as the system is cooled below this energy scale, single crystal neutron diffraction measurements were performed on CORELLI, which uses a stochastic chopper to isolate quasielastic scattering processes.

Figs.~\ref{fig:CORPLOTS} (a)-(e) show 2D slices of scattering collected in the (\textit{H}, \textit{K}, 0.25) plane with the interplane vector \textit{L} integrated from $0.15$ to $0.35$ and Figs.~\ref{fig:CORPLOTS} (f)-(j) show 2D slices of the (0.25, \textit{K}, \textit{L}) plane with the interchain vector \textit{H} integrated from $0.15$ to $0.35$. The bounds of \textit{L} and \textit{H} integration were chosen to be away from the 3D AF zone-center in order to highlight the quasi-1D diffuse nature of the magnetic signal. Data collected at 300 K show no spin correlations evident about the $K=\pm 0.5$ quasi-1D zone centers, and upon cooling to $T=100$ K magnetic correlations emerge as planes of diffuse scattering.  This diffuse, quasi-1D signal intensifies as the temperature is decreased toward 60 K. We note here that each panel taken below 300 K in Fig.~\ref{fig:CORPLOTS} also reveals maxima located within the planes of magnetic scattering. These maxima are due to the intersection of diffuse rods cutting through the $(H, K, 0.25)$ and $(0.25, K, L)$ planes and are driven by structural faulting of scattering by magnetic twins as discussed in Sect.~\ref{sec:structure}.

As the material is cooled below 60 K, the quasi-1D planes of magnetic scattering begin to weaken and are strongly suppressed upon entering the AF ordered state.  This is shown in Figs.~\ref{fig:CORPLOTS} (d), (i) and (e), (j) where the scattering centered at the 1D magnetic zone center nearly vanishes at 5 K.  Notably however, the diffuse scatter does not completely vanish at 5 K and presages the effects of microstructure and critical fluctuations in disrupting the coherence of the ordered state. 

\subsection{\label{sec:IC}Incommensurate modulated state: $T_{\mathrm{IC}}<T<T_{\mathrm{N}}$}
Upon cooling below 60 K, interchain correlation lengths grow, and below $T_{\mathrm{IC}}=45$ K a region of magnetic coexistence emerges. In this regime, an incommensurate peak appears in addition to the quasi-1D correlations centered at the $K=0.5$ position. Figs.~\ref{fig:meshfits} (a)-(c) (taken with open$-80^\prime-80^\prime-$open collimation) show representative intensity maps collected via high resolution triple-axis measurements at temperatures above ($50$ K), within ($30$ K), and below ($5$ K) this first order regime. Immediately evident from the 30 K data in Fig.~\ref{fig:meshfits} (b) is the formation of an incommensurate peak displaced along $K$ from the $(0.5, 0.5, 0)$ position that also reflects anisotropic, quasi-1D, spin correlations. The higher zone, $(1.5, 0.5, 0)$, also revealed this incommensurate peak, with the same displacement from the commensurate position. Additionally, a check of the $(0.5, 0.5, -1)$ zone showed the incommensurate peak, although in this $(H,H,L)$ scattering geometry, the displacement along $H$ versus $K$ could not be determined definitively. 

\begin{figure}[!t]
	\includegraphics[scale=0.4]{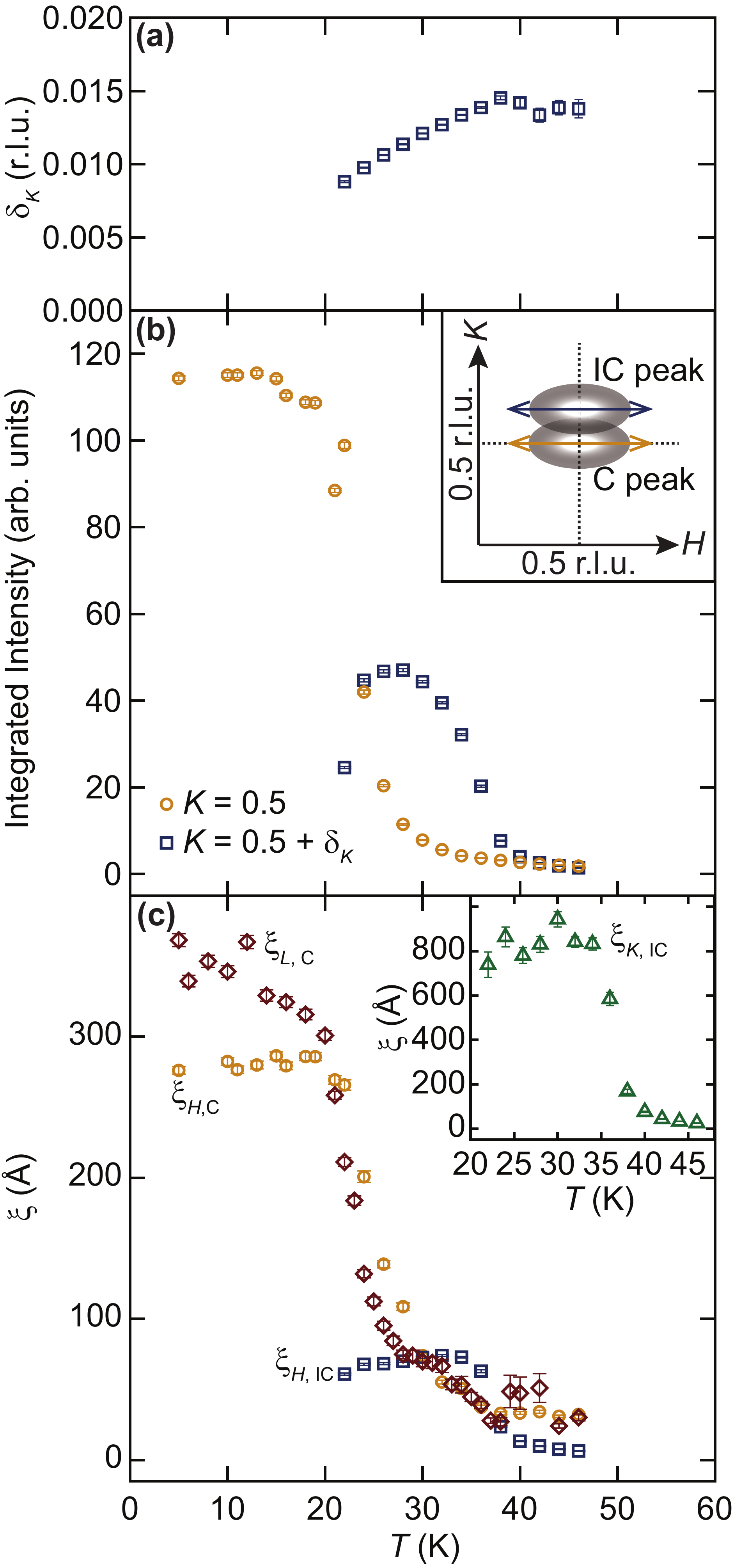}
	\caption{\label{fig:AreaCorr}(Color online) Temperature dependent parameters extracted by fitting $H$, $K$, and $L$ scans at the commensurate (C) and/or incommensurate (IC) Bragg peak positions. Panel (a) shows the center positions of the incommensurate peak, with respect to the commensurate position, along \textit{K}, where each data point has a different fixed $H=H_{\mathrm{IC}}(T)$, which is the center of the incommensurate peak along \textit{H}. (b) The integrated area of $H$ scans taken either at $K=K_{\mathrm{C}}=0.5$ r.l.u. (orange circles) or at $K=K_{\mathrm{IC}}(T)$ (blue squares). The inset is a schematic showing the direction and location of each scan. (c) The correlation length along $H$ for the C ($\xi _{H, \mathrm{C}}$) and IC ($\xi _{H, \mathrm{IC}}$) peaks, as well as along $L$ for the C peak ($\xi _{L, \mathrm{C}}$), as a function of temperature. The inset shows the correlation lengths along $K$ of the IC peak, $\xi _{K, \mathrm{IC}}$, as a function of temperature, where $H=H_{\mathrm{IC}}(T)$.}
\end{figure}

Figs.~\ref{fig:meshfits} (d)-(g) show representative line scans along both $H$ and $K$ at select temperatures, where solid black lines are the Voigt function fits to the peaks. Two peaks along $K$ are clearly visible in the insets of Figs.~\ref{fig:meshfits} (e)-(f), and the centers of these peaks are denoted as $K_{\mathrm{C}}$ and $K_{\mathrm{IC}}$, corresponding to commensurate (C) and incommensurate (IC) peaks, respectively. The corresponding $H$ scans in the main panels of Figs.~\ref{fig:meshfits} (e)-(f) were collected through the $(0.5,0.5,0)$ and $(0.5,K_{\mathrm{IC}},0)$ positions. We note that while some slight incommensurability along $H$ towards lower momentum is observed for the IC peak, this shift in the center position is not well-defined.  Namely, the intensity distribution along the interchain direction is highly diffuse and the corresponding spin-spin correlation lengths (shown in Fig.~\ref{fig:AreaCorr} (c)) at the onset of IC order are substantially smaller than the long-wavelength modulation implied to the apparent incommensurability.  The apparent shift (discussed further in Sect.~\ref{sec:disc}) is interpreted to be extrinsic in origin and likely driven by the resolution of the triple-axis spectrometer convolving out-of-plane scattering components into the $(H, K, 0)$ plane. 

The $T_{\mathrm{IC}}=45$ K transition temperature was extracted via the highest temperature where an incommensurate peak was resolvable in $K$-scans through the $(0.5, 0.5, 0)$ position.  As the material is cooled below $45$ K, the incommensurate peak position $(0.5, 0.5+\delta _{K}(T), 0)$ shifts below $\delta _K =0.015$ toward the commensurate $\delta _K =0$ wave vector as parametrized in Fig.~\ref{fig:AreaCorr} (a). As the IC peak shifts, it grows in intensity and sharpens, and the temperature dependence of integrated intensities as well as correlation lengths are plotted in Figs.~\ref{fig:AreaCorr} (b) and (c). For comparison, the temperature dependence of the integrated intensities and correlation lengths of the commensurate order are overplotted in Figs.~\ref{fig:AreaCorr} (b) and (c).

To further characterize the spin order in this regime, $L$ scans through the $(0.5, 0.5, 0)$ position were also collected. Interplane correlation lengths of the commensurate peak $\xi_{L\mathrm{, C}}$ were found to be similar to the interchain lengths $\xi_{H\mathrm{, C}}$ as shown in Fig.~\ref{fig:AreaCorr} (c). Fig.~\ref{fig:AreaCorr} (c) shows that the IC peak also remains quasi-1D, and, like the commensurate peak, has a substantially longer correlation length $\xi_{K\mathrm{, IC}}$ (intrachain) relative to $\xi_{H\mathrm{, IC}}$ (interchain). The correlation length along $K$ for the commensurate peak is not plotted as it rapidly reaches the limit of resolution ($>1500$ \AA) around $T=35$ K. 

After the initial growth of its spectral weight and an increase in its correlation lengths, the IC peak saturates at 33 K. Below this temperature, the IC peak decreases in intensity and ultimately vanishes/convolves into the commensurate position below 22 K.  Simultaneous to the disappearance of the IC peak, the intensity and in-plane correlation lengths of the $(0.5, 0.5, 0)$ N{\'e}el state saturate. This demonstrates an interplay of the two coexisting states in which the IC modulated spin state slowly condenses into the expected N{\'e}el state below 22 K.

\subsection{N{\'e}el phase: $T<T_{\mathrm{N}}\approx 22$ K}

Below $T_{\mathrm{N}}=22$ K, a single commensurate peak can be fit to $K$-scans through $(0.5, 0.5, 0)$.  Measurements deep in the ordered state at $T = 5$ K reveal an incomplete transition into the 3D-ordered state with finite correlation lengths along $H$ and $L$ shown in Fig. 5.  This quasi-1D anisotropy can also be visualized in the 5 K intensity map of Fig.~\ref{fig:meshfits} (a) and the corresponding $H$ and $K$ line-scans in Fig.~\ref{fig:meshfits} (d). While the sharp exchange of intensity at $T_\mathrm{N}$ between the commensurate and incommensurate order parameters implies the modulated moments of the IC state lock into the commensurate position, structural disorder effects disrupt the formation of a truly long-range ordered N{\'e}el state. 

\begin{figure}[!t]
	\includegraphics[scale=0.4]{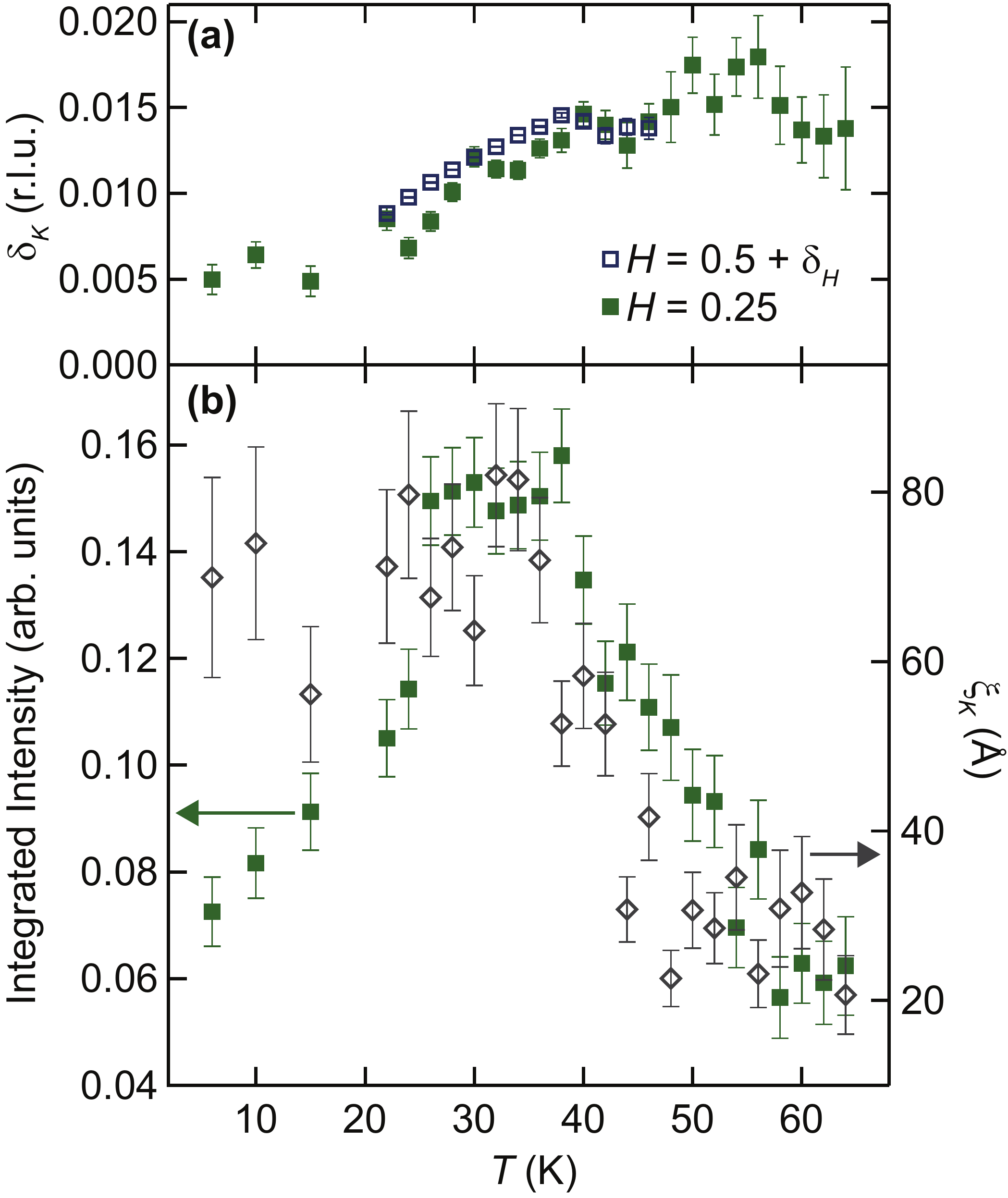}
	\caption{\label{fig:KcenterZB}(Color online) Tracking of the incommensurate scattering at the magnetic zone boundary as a function of temperature. The green filled squares in (a) reflect the tracking of the diffuse critical scattering at the zone boundary and represent the peak of the diffuse scattering along \textit{K} at the zone boundary (ZB), where \textit{H}$=0.25$ r.l.u.. The open blue squares are the same data from Fig.~\ref{fig:AreaCorr} (a) near the magnetic zone center. The filled green squares in panel (b) show the integrated area of the diffuse scattering peak at the ZB, and the open gray diamonds represent the correlation length of the diffuse scattering at the ZB. These data were taken with open$-80'-80'-$open collimation.}
\end{figure}

Diffuse scattering shown in Figs.~\ref{fig:KcenterZB} (a) and (b) associated with the IC, quasi-1D, wave vector and measured at the 3D magnetic zone boundary (0.25, $0.5+\delta_{K}(T)$, 0) also remains resolvable below 22 K.  Intrachain correlation lengths $\xi_K$ for this quasi-1D diffuse component sharpen and saturate in intensity at 22 K suggesting that these quasi-1D correlations are critical fluctuations driving the onset of N{\'e}el order.  This weaker 1D critical scattering underlies and tracks the IC peak's position as it approaches the zone center, as evidenced by the data from Fig.~\ref{fig:AreaCorr} (a) overplotted with the zone boundary data in Fig.~\ref{fig:KcenterZB} (a). The fact that critical fluctuations remain robust at 5 K suggest that the transition into long-range AF order is incomplete.

\section{\label{sec:disc}Discussion}

Previous powder neutron measurements were able to resolve the formation of short-range spin correlations at high temperatures as well as their persistence within the ordered state of NaMnO$_2$; however the dimensionality of these fluctuations and their interplay with the formation of N{\'e}el order remained ambiguous.\cite{Giot_2007} Our single crystal data are now able to probe the details of the onset of short-range order and its evolution into the ordered phase. Specifically, the reciprocal space map collected at 100 K establish that the high temperature correlations as one dimensional in character, and upon cooling, these fluctuations seemingly drive the formation of two competing short-range ordered states---the N{\'e}el state and an incommensurately modulated spin structure. 

Many systems with similar lattice topologies are known to host intermediate, incommensurately modulated spin textures, which subsequently transition into a collinear antiferromagnetic state upon cooling into the ground state. Li$_2$NiW$_2$O$_8$,\cite{ Ranjith_PRB_2016} CoNb$_2$O$_6$,\cite{Columbite_JMMM_1995, Columbite_PRB_1999}, Cs$_2$CuCl$_4$\cite{Coldea_JPhysCM_1996, Isakov_PRB_2005} and CuFeO$_2$\cite{Mitsuda_JPSJ_1998, Petrenko_JPhysCM_2005, Ye_PRB_2006_CuFeO2} are all examples of quasi-1D materials exhibiting this behavior. Here CuFeO$_2$ is particularly germane to $\alpha$-NaMnO$_2$, as it shares a triangular lattice whose lattice symmetry is lowered into $C2/m$ symmetry upon cooling. This generates an anisotropic triangular lattice of Cu$^{2+}$ moments where an incommensurate, sinusoidally modulated spin configuration sets in with a temperature dependent wave vector. Similar to $\alpha$-NaMnO$_2$, the modulation of moments eventually locks into the commensurate wave vector and true long-range order fails to stabilize. 

In $\alpha$-NaMnO$_2$, the observation of a satellite peak about the commensurate $(0.5, 0.5, 0)$ position is suggestive of modulated magnetic order. However, the $C2/m$ space group with the single magnetic site at $(0, 0, 0)$ leads to only one independent magnetic sublattice, and coupled with the magnetic propagation wave vector, $\mathbf{k}_{\mathrm{IC}} = (0.5, 0.5+\delta_K, 0)$ (or, alternatively, $\mathbf{k}_{\mathrm{IC}} = (0.5+\delta_H, 0.5+\delta_K, 0)$), there are no irreducible representations that result in extinction rules which reflect the inequality between pairs of satellite peaks  (i.e.\ intensity at $\mathbf{Q}=(1,1,0)-\mathbf{k}_{\mathrm{IC}}$ versus intensity at $\mathbf{Q}=(0,0,0)+\mathbf{k}_{\mathrm{IC}}$). With the exception of differences derived from the magnetic form and orientation factors, the magnetic structure factor $\mathbf{F}_M \left( \mathbf{Q} \right)$ is equal for all $\mathbf{Q}$. The data presented in Sec.~\ref{sec:IC} indicate complete extinction, or else a very small structure factor, for $\mathbf{Q}=(1,1,0)-\mathbf{k}_{\mathrm{IC}}$ and a much larger structure factor for $\mathbf{Q}=(0,0,0)+\mathbf{k}_{\mathrm{IC}}$. Strong asymmetry in the structure factors for pairs of satellite peaks can only occur if there is a lowering of the crystallographic symmetry such that more than one magnetic sublattice and basis vector mode can be defined (such as in Ref.~\onlinecite{Coldea_JPhysCM_1996}). Several prior studies\cite{Ouyang_PRB_2010, Zorko_PRB_2008, Giot_2007, Jia_JAppPhys_2011} of $\alpha$-NaMnO$_2$ powders have suggested that spin-lattice coupling drives a weak structural distortion into a lower symmetry, triclinic cell. Although the resolution of our current single crystal experiments was insufficient to resolve this structural change, a similar distortion is likely present in our crystals and can allow for the formation of a modulated state where the intensity between pairs of incommensurate satellite peaks is unequal. 

\begin{figure}[t]
\includegraphics[scale=0.40]{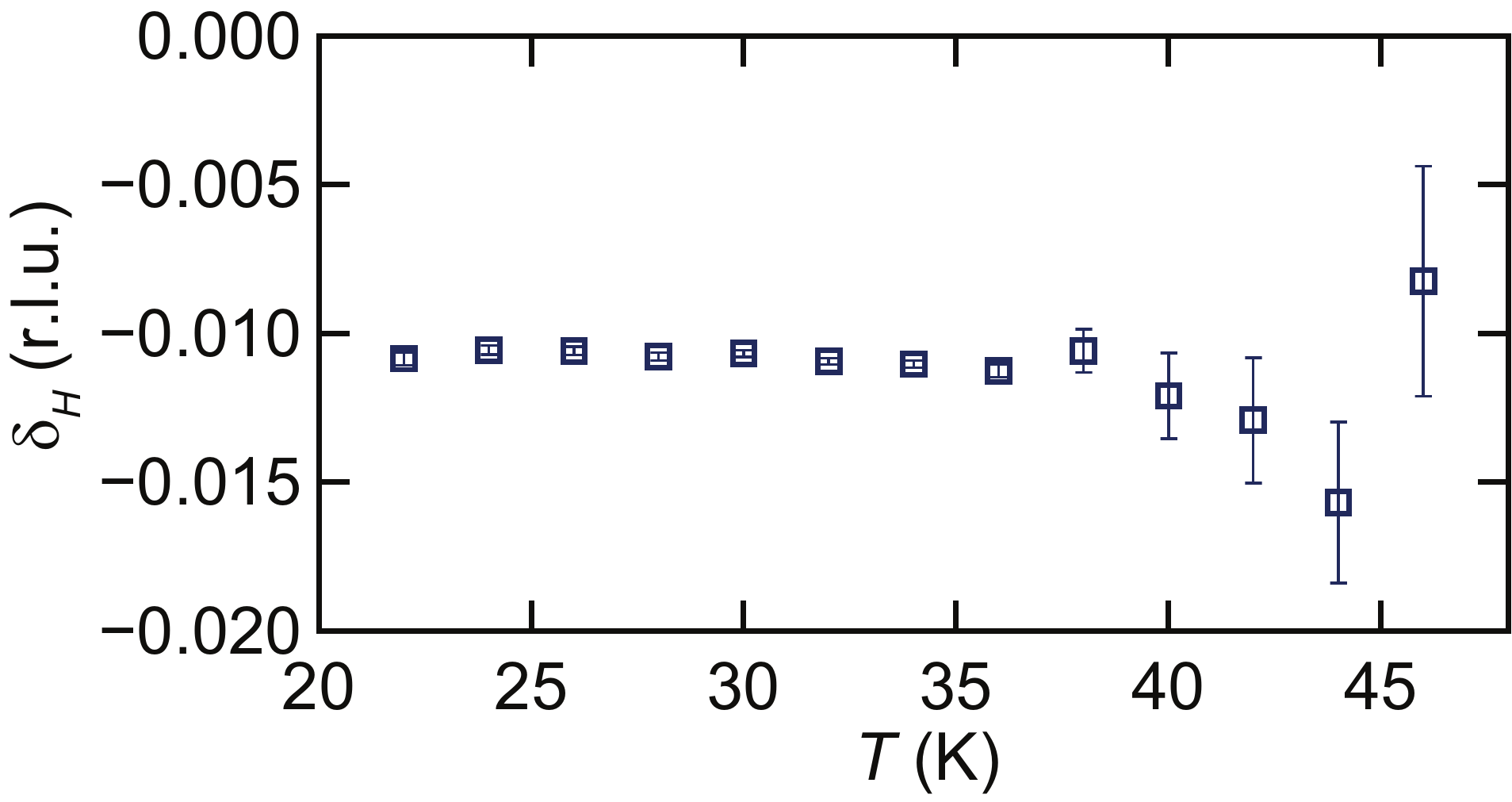}
\caption{\label{fig:Hcenter}(Color online) The center position of the incommensurate peak, with respect to the commensurate position, along \textit{H}. Each data point has a different fixed $K=0.5+\delta _K (T)$, which is the center of the incommensurate peak along \textit{K}.}
\end{figure}

As mentioned in Sect.~\ref{sec:IC}, there also exists an apparent incommensurability along $H$ for the IC peak, which is summarized in Fig.~\ref{fig:Hcenter}. This offset is small and does not follow the trend of $\delta_K$ in sliding toward the commensurate position. While at 45 K an incommensurability of $\approx(-\delta_H,\delta_K,0)$ would be consistent with a displacement along one of the principle axis of the lower symmetry triclinic unit cell, \cite{Giot_2007} its rotation with temperature (due to temperature independent $\delta_H$) would require further high resolution structure measurements to connect the two phenomena. Additionally, the correlation length along $H$ for the incommensurate peak at its onset is only $\approx 7$ {\AA}, far below the modulation wavelength implied by the wave vector. 

Rather, a more likely possibility is that the apparent incommensurability along $H$ is an artifact of the resolution function of the spectrometer convolving the convergence of both quasi-1D magnetic features and the quasi-2D rod along $[\overline{1}, 0, 1]$ close to the magnetic zone center. Modeling the two quasi-1D commensurate and incommensurate peaks with ResLib using the Cooper-Nathans approximation\cite{Cooper_1967} can produce a slight shift of the incommensurate peak towards lower-Q in a simulated $H$ scan of the IC peak, however the magnitude of the simulated shift was insufficient to unambiguously explain the experimental observation.  This implies that an out-of-plane component such as the structurally driven magnetic rods of scattering also acts to bias the peak position.

If one assumes the intrinsic incommensurate modulation of moments is only along the chain direction, $K$, our data can be compared with prior mean-field models of Ising-like moments on an anisotropic triangular lattice. Specifically, the dominant intrachain $J_1$ and weaker interchain $J_2$ exchange couplings can be used to predict the modulation wave vector at the onset of the incommensurate phase in the mean field approximation via $\delta _K=\frac{1}{\pi }\arcsin (\frac{J_2}{2J_1})$.\cite{Dalidovich_PRB_2006, Columbite_PRB_1999}  Using values for $\alpha$-NaMnO$_2$ from Ref.~\onlinecite{Dally_NatComm}, this mean-field model predicts a $\delta _K =0.02$ r.l.u., which is close to the maximum resolvable displacement of $\delta _K (T=45$ K$)=0.015$ r.l.u. observed in experiments.

A notable difference between $\alpha$-NaMnO$_2$ and many other anisotropic triangular lattice compounds is that within the temperature regime just above $T_{\mathrm{N}}$ there is a coexistence of incommensurate short-range order and commensurate short-range magnetic states. One interpretation of the simultaneous presence of critical scattering and a first-order order coexistence between the incommensurate and commensurate phases could be the influence of a nearby Lifshitz multicritical point.\cite{Sinha_PRB_1981, hornreich1980lifshitz} However, a more likely explanation for this coexistence is rooted in the structural complexity of $\alpha$-NaMnO$_2$.  In-depth studies \cite{Zorko_NatComm_2014, Zorko_SciRep_2015} on the structural inhomogeneity and magnetoelastic effects within $\alpha$-NaMnO$_2$ powders report that the system does not undergo a bulk transition into a triclinic phase, but instead is phase separated into nanoscale regions. The coexistence of magnetic phases can therefore arise as a consequence of the structural inhomogeneity hypothesized to be inherent to this system. A similar coexistence of phases is also reported in the structurally similar material NaFeO$_2$. \cite{McQueen_2007} 

Despite the influence of structural disorder through twinning and stacking faults, the study of $\alpha$-NaMnO$_2$ crystals provides the first evidence for an intermediate incommensurate magnetic state that stabilizes prior to the onset of previously reported N\'{e}el order. Access to anisotropies in momentum space furthermore indicate that spin correlations in the N\'{e}el phase can remain quasi-1D down to 5 K, likely due to microstructure effects. This behavior is consistent with the dominant one dimensional exchange of the material with an anisotropy ratio $J_2/J_1=0.12$---among the most anisotropic of reported planar triangular lattices. An intriguing possibility is that the quasi-1D, quasielastic scattering observed deep in the AF ordered phase of $\alpha$-NaMnO$_2$ is driven by the spin amplitude mode recently reported in this system. \cite{Dally_NatComm}  Such a mode should serve to destabilize long-range AF order.

\section{Conclusions}

$\alpha$-NaMnO$_2$ is inherently a quasi-1D magnet where intrachain magnetic correlations develop at temperatures as high as 200 K. Upon cooling below this energy scale, the Mn$^{3+}$ moments within the spatially anisotropic triangular lattice host coexisting commensurate and incommensurate short-range ordered states between $T_{\mathrm{IC}}=45$ K$<T< T_{\mathrm{N}}=22$ K. The quasi-1D incommensurate phase is characterized via a temperature dependent wave vector modulated along the chain axis that slides towards the commensurate N\'{e}el position upon cooling.  Below $T_{\mathrm{N}}=22$ K, the incommensurate order merges with or ``locks" into the commensurate state, where anisotropic correlation lengths develop due to the crystal microstructure. The intermediate, lock-in phase behavior is consistent with the expectations of a mean-field model of classical Ising spins on a highly anisotropic triangular lattice, and the richness of the magnetic phase behavior within $\alpha$-NaMnO$_2$ motivates future studies to further explore the behavior of this quasi-1D spin system.

\appendix* 
\section{Intergrowth of Mn$_3$O$_4$}
\subsection{Determining the orientation and molar fraction}
\begin{figure}[t]
\includegraphics[scale=0.64]{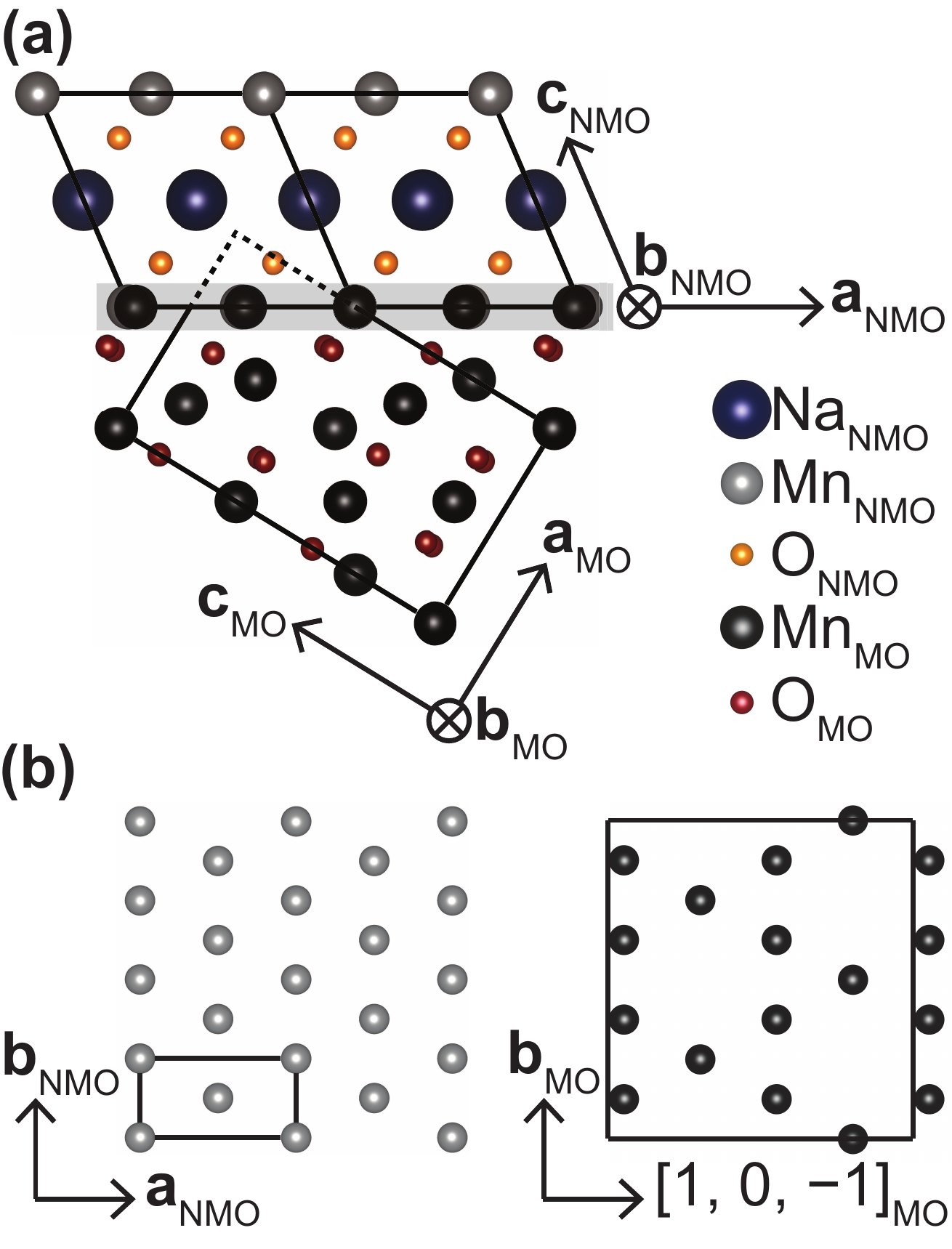}
\caption{\label{fig:intergrowth}(Color online) Schematic of the Mn$_3$O$_4$ (MO) intergrowth with $\alpha$-NaMnO$_2$ (NMO). (a) The $ac$-plane of both phases shown meeting at the boundary of intergrowth, highlighted in gray. The chemical unit cells of each phase are outlined with solid black lines, and the extension of the MO unit cell into the NMO unit cell is depicted as a dashed black line. (b) Comparison of two parallel planes: the $ab$-plane of $\alpha$-NaMnO$_2$ and the Mn$_3$O$_4$ plane spanned by $\mathbf{b}_{\mathrm{MO}}$ and $[1, 0, -1]_{\mathrm{MO}}$.}
\end{figure}

A small impurity phase of Mn$_3$O$_4$ was found intergrown within the single crystals of $\alpha$-NaMnO$_2$ studied here. It is suspected that during floating zone growth, Na diffuses out of the crystal near domain boundaries to form Na-deficient regions that stabilize into Mn$_3$O$_4$.\cite{MOphases} Since the atom-atom pair correlations for Mn$_3$O$_4$ are a close subset of those within $\alpha$-NaMnO$_2$, detection of this impurity is not easily done via powder diffraction of crushed single crystals. However in single crystal diffraction, the Mn$_3$O$_4$ intergrowth can be resolved to have a well-defined orientation within the $\alpha$-NaMnO$_2$ lattice.  This was detectable in our reciprocal space maps from single crystal measurements, and the intergrowth orientation could be determined. The relative orientation between the regions of Mn$_3$O$_4$ intergrowth and the host $\alpha$-NaMnO$_2$ lattice is illustrated in Fig.~\ref{fig:intergrowth}. 

The boundaries between the two phases are reminiscent of the way in which twin boundaries and stacking faults are formed in $\alpha$-NaMnO$_2$.  Fig.~\ref{fig:intergrowth} (a) shows this boundary, and it can be seen that the elongated direction of the Jahn-Teller distortions in both the $\alpha$-NaMnO$_2$ and Mn$_3$O$_4$ are continuous across the junction of the two structures. The oxygen atoms do not have to move to accommodate the change in phase. It should be noted that the Mn$_3$O$_4$ phase is slightly strained from it's bulk structure. For example, the $(0, 2, 0)_{\mathrm{MO}}$ nuclear Bragg peak appears exactly at the $(0,1,0)_{\mathrm{NMO}}$ position, indicating that $\mathbf{b}_{\mathrm{MO}}=2\mathbf{b}_{\mathrm{NMO}}$, although the reported $b$-axis lattice parameter for Mn$_3$O$_4$ is 5.71 \AA \  and the refined room temperature $b$-axis lattice parameter for $\alpha$-NaMnO$_2$ is $2.8602 \pm 0.0002$ \AA. 

While this intergrowth is resolvable in diffraction data, due to the structural complexity of the host $\alpha$-NaMnO$_2$ lattice, it is difficult to reliably and quantitatively extract the impurity fraction of Mn$_3$O$_4$ from the diffraction data alone. Instead, analysis of bulk magnetization measurements facilitates the most reliable estimates of the relative fraction of the Mn$_3$O$_4$ intergrowth. The collinear antiferromagnetic ground state of $\alpha$-NaMnO$_2$ should not contribute any net magnetization, magnetic hysteresis or coercivity.\cite{Giot_2007,Zorko_PRB_2008} Therefore, within crystals containing an Mn$_3$O$_4$ impurity, any observation of weak ferromagnetism can be assumed to come from Mn$_3$O$_4$. The ferromagnetic component of the ordered moment in Mn$_3$O$_4$ single crystals of 1.89 $\mu_{\mathrm{B}}$ per formula unit determined from neutron diffraction experiments could be used;\cite{CHARDON1986} however, magnetization measurements under finite field measure a slightly lower polarized moment.\cite{Kemei_2014, Dwight_1960, Nielsen_1976} 

\begin{figure}[t]
\includegraphics[scale=0.45]{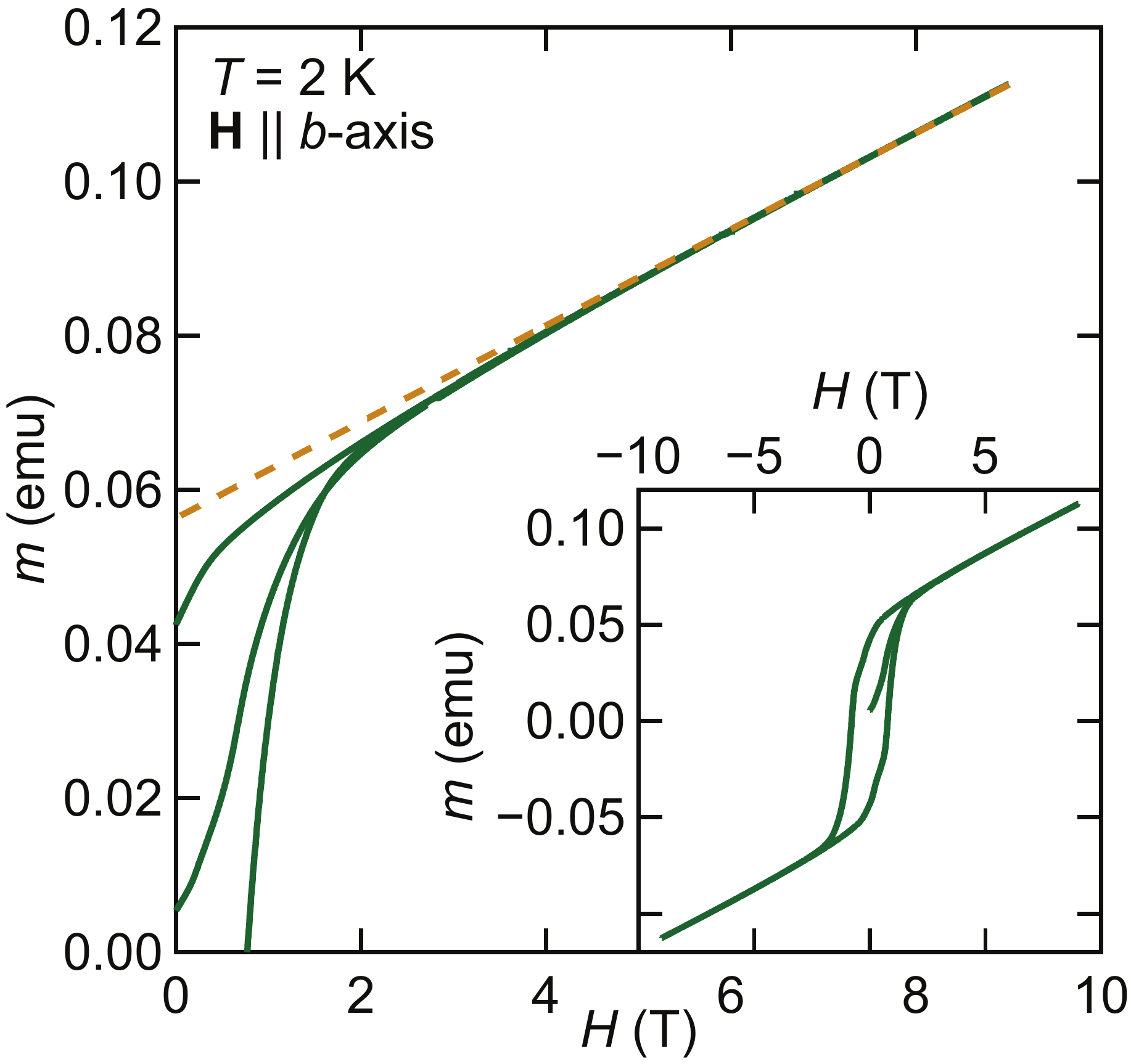}
\caption{\label{fig:mag}(Color online) Bulk magnetization measurements taken on single crystals of $\alpha$-NaMnO$_2$. The data show $T=2$ K magnetic field sweeps with $\mathbf{H} \parallel b$-axis of $\alpha$-NaMnO$_2$ and Mn$_3$O$_4$. The dashed orange line is a fit to the high field data between $8.5 < H (\mathrm{T}) < 9$ used for extracting the mass percentage of Mn$_3$O$_4$ intergrowth in the sample as described in the text.}
\end{figure}

As an upper bound of the Mn$_3$O$_4$ content we use this lower magnetization-derived value, and Ref.~\onlinecite{Suzuki_PRB_Mn3O4_SC_magnetization} reports the magnetization of Mn$_3$O$_4$ single crystals with a field oriented parallel to the easy-axis, which is the $[0, 1, 0]$ or $[1, 0, 0]$ crystallographic direction of Mn$_3$O$_4$ (note: Ref.~\onlinecite{Suzuki_PRB_Mn3O4_SC_magnetization} uses a different unit cell than here). By orienting the field along the $b$-axis of $\alpha$-NaMnO$_2$, we also align along the $[0, 1, 0]$ easy-axis of Mn$_3$O$_4$, as the $b$-axes for the two phases are parallel. In using the reported value of 1.7 $\mu_{\mathrm{B}}$ per Mn$_3$O$_4$ as the saturated moment, we can interpret the data plotted in Fig.~\ref{fig:mag} as a superposition of a linear magnetization arising from $\alpha$-NaMnO$_2$ and a saturating magnetization arising from Mn$_3$O$_4$.  A linear fit to the data (shown as the dashed orange line in Fig.~\ref{fig:mag}) between $H=8.5$ T and $H=9$ T was used to subtract the response of the collinear antiferromagnet, $\alpha$-NaMnO$_2$, and the remaining $m(H)$ at $H=9$ T was taken to be the saturated contribution from the Mn$_3$O$_4$ intergrowth. From this, the mass percentage of the intergrowth was determined to be $\approx$6\%. A key observation here is that ICP measurements of the Na:Mn molar ratio of intergrown $\alpha$-Na$_x$MnO$_2$ crystals typically yield an average Na content of $x=0.90$. Once the intergrowth of Mn$_3$O$_4$ is accounted for, the Na content in $\alpha$-NaMnO$_2$ regions is nearly stoichiometric, with $x\approx0.98$.

\subsection{\label{sec:app}Mn$_3$O$_4$ in neutron diffraction}

\begin{figure}[b]
\includegraphics[scale=0.31]{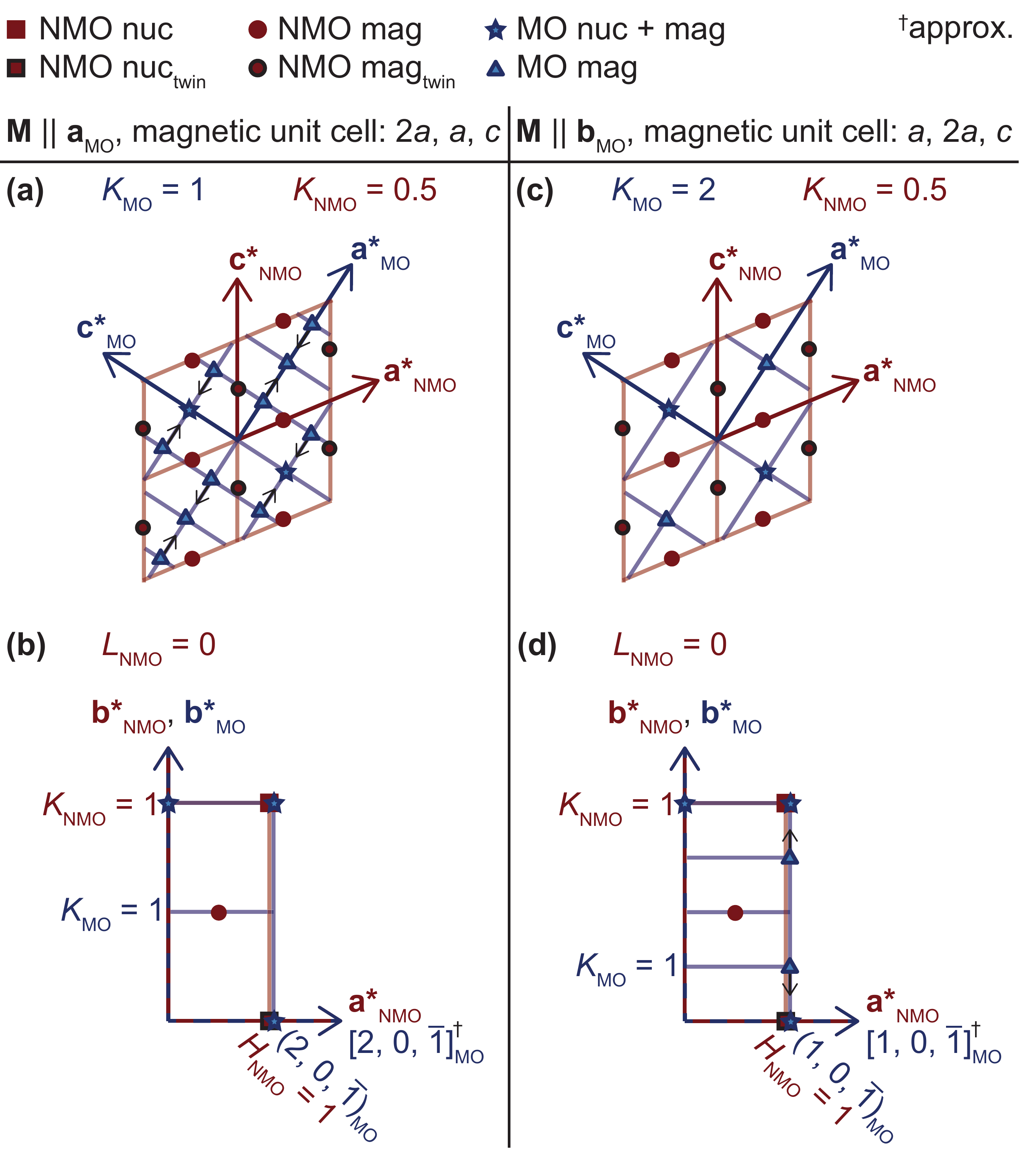}
\caption{\label{fig:intergrowthrecip}(Color online) Reciprocal space maps for the Mn$_3$O$_4$ magnetic unit cell and $\alpha$-NaMnO$_2$ chemical unit cell, orientated so that they reflect the real space intergrowth direction. Panels (a) and (b) assume the \textbf{a}-domain magnetic structure for Mn$_3$O$_4$, and (c) and (d) assume the \textbf{b}-domain magnetic structure for Mn$_3$O$_4$. Panels (a) and (c) show the $a^*c^*$-plane for both Mn$_3$O$_4$ and $\alpha$-NaMnO$_2$, where the $b^*$-axis is going into the page for both phases. Panels (b) and (d) show the $a^*b^*$-plane for $\alpha$-NaMnO$_2$ and the plane which is parallel in Mn$_3$O$_4$: either the $(2H,K,-H)_{\mathrm{MO}}$ for the \textbf{a}-domain or the $(H,K,-H)_{\mathrm{MO}}$ for the \textbf{b}-domain. Integral miller index positions for $\alpha$-NaMnO$_2$ are defined by the intersections of red lines and integral miller index positions for Mn$_3$O$_4$ are defined by the intersections of blue lines. Allowed Bragg peaks are marked according to the legend at the top of the figure. The black arrows represent the direction in which magnetic Bragg peaks shift for Mn$_3$O$_4$ when in its incommensurate phase as reported in Ref.~\onlinecite{Jensen_1974}.}
\end{figure}

Careful consideration of the Mn$_3$O$_4$ intergrowth's contribution to magnetic scattering must be made to ensure that it does not contribute to data exploring the magnetic states of $\alpha$-NaMnO$_2$. The relative positions in reciprocal space where the nuclear and magnetic Bragg reflections from the intergrown Mn$_3$O$_4$ appear are illustrated in Fig.~\ref{fig:intergrowthrecip}. As Mn$_3$O$_4$ is tetragonal, there are two magnetic $k$-domains associated with the ordering wave vector: $\mathbf{M} \parallel \mathbf{a}$ (\textbf{a}-domain) and $\mathbf{M} \parallel \mathbf{b}$ (\textbf{b}-domain). Both were taken into consideration when determining the location of potential scattering from the Mn$_3$O$_4$ intergrowth.

\begin{figure}[b]
\includegraphics[scale=0.45]{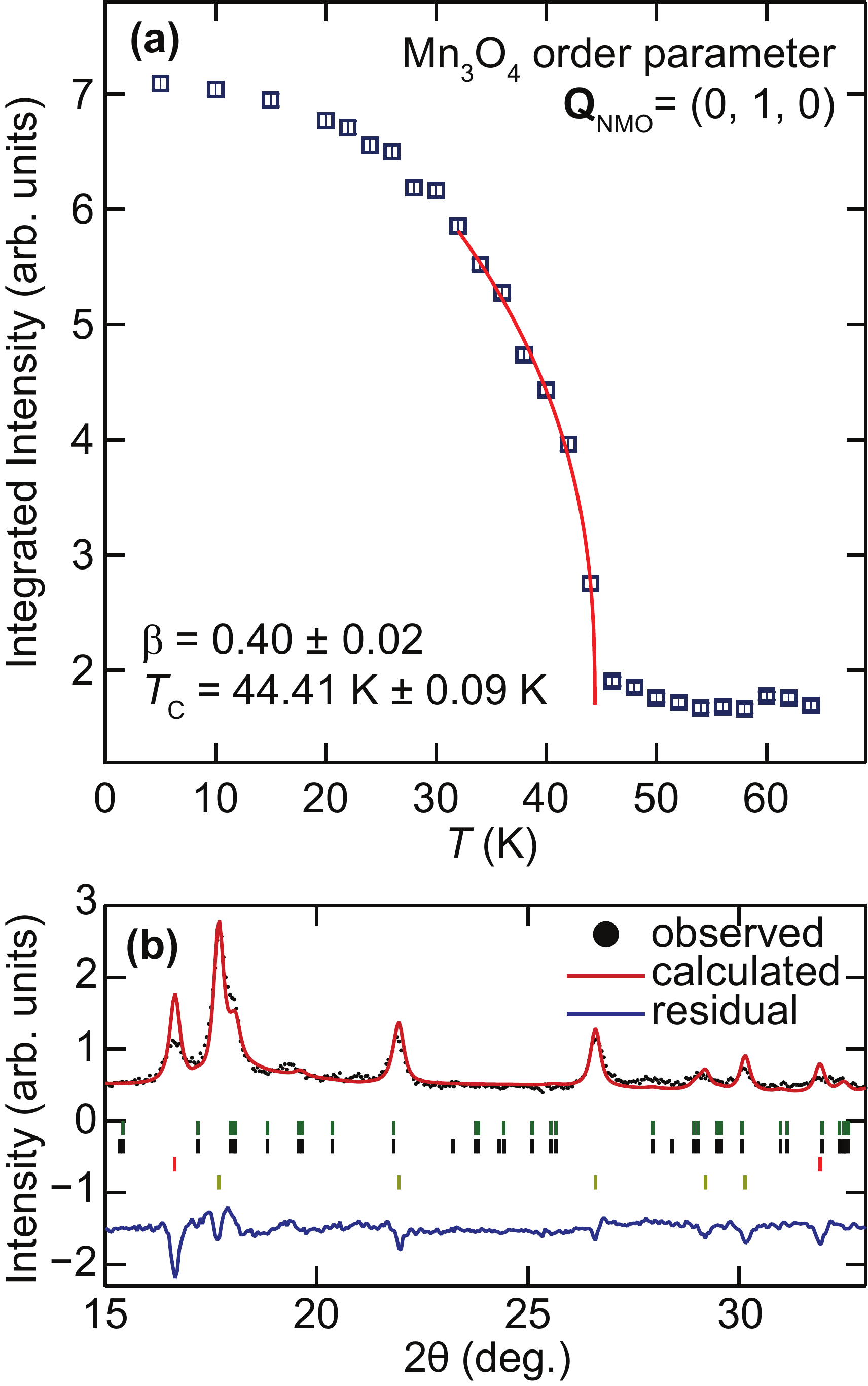}
\caption{\label{fig:magNPD}(Color online) Neutron data showing the presence of Mn$_3$O$_4$ magnetic impurity within $\alpha$-NaMnO$_2$. (a) Triple-axis neutron data of integrated \textit{K}-scans centered at $\textbf{Q}=(0, 1, 0)_{\mathrm{NMO}}$. This momentum corresponds to the $(0,2,0)_{\mathrm{MO}}$ in the \textbf{a}-domain, and $(0, 4, 0)_{\mathrm{MO}}$ in the \textbf{b}-domain, of Mn$_3$O$_4$. The solid red line is a power law fit to the data yielding the critical exponent, $\beta$, and ordering temperature, $T_{\mathrm{C}}$, as indicted. (b) Neutron powder diffraction data taken at $T=5$ K of a crushed single crystal of $\alpha$-NaMnO$_2$. The nuclear and magnetic structure refinement using the Rietveld method is shown as a solid red line. The positions of Bragg peaks are marked with ticks underneath the spectra. Going from top to bottom they represent the Mn$_3$O$_4$ nuclear, Mn$_3$O$_4$ magnetic, $\alpha$-NaMnO$_2$ nuclear, and $\alpha$-NaMnO$_2$ magnetic Bragg peak positions.}
\end{figure} 

To ensure the Mn$_3$O$_4$ intergrowth behaves similar to the bulk, the temperature dependence of the order parameter at Mn$_3$O$_4$ (0, 4, 0) magnetic Bragg peak (using notation for the \textbf{b}-domain) was collected. There should be no nuclear or magnetic contribution from the $\alpha$-NaMnO$_2$ at this wave vector (corresponding to the (0, 1, 0) of $\alpha$-NaMnO$_2$), and the resulting order parameter is shown in Fig.~\ref{fig:magNPD} (a). A power law fit to the form $I\propto((T_{\mathrm{C}}-T)/T_{\mathrm{C}})^\beta$ yields $\beta=0.40 \pm 0.02$ and a critical temperature, $T_{\mathrm{C}}=44.41$ K$ \pm 0.09$ K. While Mn$_3$O$_4$ is known to undergo multiple phase transitions below this temperature, starting from a reported $T_{\mathrm{C}}=41$ K,\cite{CHARDON1986, Jensen_1974} we are able to ensure that the incommensurate scattering attributed to $\alpha$-NaMnO$_2$ does not arise from those phases. Near the $\alpha$-NaMnO$_2$ zone center, there are no structural or magnetic Mn$_3$O$_4$ peaks allowed and none close enough to interfere with the study of the incommensurate phase endemic to $\alpha$-NaMnO$_2$, as shown in Fig.~\ref{fig:intergrowthrecip}.  

As a rough consistency check of the relative abundance of the structural and magnetic states of Mn$_3$O$_4$ within $\alpha$-NaMnO$_2$ crystals, low temperature neutron powder diffraction data were also collected on a crushed single crystal. Low angle diffraction data taken at $T=5$ K are shown in Fig.~\ref{fig:magNPD} (b) where the refinement includes a structural model incorporating both the $\alpha $-NaMnO$_2$ and Mn$_3$O$_4$ lattices (in the ratio determined via magnetization measurements), as well both of their magnetic phases. The Mn$_3$O$_4$ magnetic phase was assumed to be published values and was not refined. \cite{Jensen_1974}  We note that due to the absence of stacking faults in the model, there will be a large mismatch between the calculated and measured $\alpha $-NaMnO$_2$ structural peak intensities as discussed in Sect.\ III.  

The Mn moments from $\alpha$-NaMnO$_2$ were refined using irreducible representational analysis with the basis vectors presented in Table~\ref{tab:basisvectors}.  Only basis vectors $\bfpsi_{1}$ and $\bfpsi_{3}$, with coefficients of opposite sign, were needed to capture the magnetic peak intensities, and attempts to refine $\bfpsi_{2}$, which represents a projection of the moment along the $b$-axis, refined to be zero within error. The collinear Mn moment orientation refined to orient $38(2)^{\circ}$ out of the $ab$-plane (and perpendicular to the $b$-axis), with the projections along the \textit{a} and \textit{c} axes $m(a)=-1.29(7)$ $\mu _{\mathrm{B}}$ and $m(c)=1.61(7)$ $\mu _{\mathrm{B}}$, respectively. The total saturated moment of $2.42(7)$ $\mu _{\mathrm{B}}/\mathrm{Mn}^{3+}$ is significantly reduced from the full spin-only value of $4$ $\mu _{\mathrm{B}}/\mathrm{Mn}^{3+}$, as well as the previous observation in powders of 2.92 $\mu _{\mathrm{B}}/\mathrm{Mn}^{3+}$. \cite{Giot_2007} The reduced magnitude relative to the prior powder result likely stems from the current analysis's failure to account for structural faulting (as evidenced by the rather poor fit quality in this rough consistency check). Nevertheless, this crosscheck confirms the relative abundance of Mn$_3$O$_4$ intergrown within $\alpha$-NaMnO$_2$ crystals as well as our models of its magnetic contributions. 

\begin{table}[t]
	\caption[Basis vectors of the irreducible representation for the collinear antiferromagnetic ground state in $\alpha$-NaMnO$_2$]{Basis vectors for the space group $C2/m$ with ${\bf k} =( 0.5,~ 0.5,~ 0)$. The decomposition of the magnetic representation for the Mn site 
		$( 0,~ 0,~ 0)$ is $\Gamma_{Mag}=3\Gamma_{1}^{1}+0\Gamma_{2}^{1}$. Table output was created using SARAh \cite{SARAh}.}
	\label{tab:basisvectors}
	\begin{tabular}{ccc|cccccc}
		IR  &  BV  &  Atom & \multicolumn{6}{c}{BV components}\\
		&      &             &$m_{\|a}$ & $m_{\|b}$ & $m_{\|c}$ &$im_{\|a}$ & $im_{\|b}$ & $im_{\|c}$ \\
		\hline
		$\Gamma_{1}$ & $\bfpsi_{1}$ &      1 &      2 &      0 &      0 &      0 &      0 &      0  \\
		& $\bfpsi_{2}$ &      1 &      0 &      2 &      0 &      0 &      0 &      0  \\
		& $\bfpsi_{3}$ &      1 &      0 &      0 &      2 &      0 &      0 &      0  \\
	\end{tabular}
\end{table}

\begin{acknowledgments}

S.D.W. and R.L.D. gratefully acknowledge support from DOE, Office of Science, Basic Energy Sciences under Award DE-SC0017752. The research reported here made use of shared facilities of the UCSB MRSEC (NSF DMR 1720256), a member of the Materials Research Facilities Network (www.mrfn.org). A portion of this research used resources at the Spallation Neutron Source, a DOE Office of Science User Facility operated by the Oak Ridge National Laboratory. We acknowledge the support of the National Institute of Standards and Technology, U.S. Department of Commerce, in providing the neutron research facilities used in portions of this work.  S.D.W. and R.L.D. also acknowledge support from the California NanoSystems Institute (CNSI) 
\end{acknowledgments}

\end{document}